\documentclass[twocolumn,showpacs,preprintnumbers,amsmath,amssymb,floatfix]{revtex4}
\usepackage[applemac]{inputenc}
\usepackage[T1]{fontenc} 
\usepackage{subfigure}
\usepackage{amsmath}
\usepackage{amsfonts}
\usepackage{amssymb}
\usepackage[dvips]{graphicx}
\usepackage[normalem]{ulem}
\usepackage[colorlinks]{hyperref}
\usepackage[usenames,dvipsnames]{color}

\usepackage{bbm}
\def\beq{\begin{equation}}
\def\eeq{\end{equation}}
\def\bea{\begin{eqnarray}}
\def\eea{\end{eqnarray}}
\def\ba{\begin{array}}
\def\ea{\end{array}}

\def\part{\partial}

\def\ie{{\it{i.e. }}}

\begin{document}

\preprint{UdeM-GPP-TH-23-300}

\title{Thin-Wall Monopoles in a False Vacuum}
\author{M. B. Paranjape$^1$}
\email{paranj@lps.umontreal.ca}
\author{Yash Saxena$^{2}$}
\email{1980234@pg.du.ac.in}

\affiliation{$^1$Groupe de physique des particules, Département de
physique, Centre de recherche mathématiques and Institut Courtois, Universit\'e de Montr\'eal, C.P. 6128, Succ. Centreville, Montr\'eal, Qu\'ebec, CANADA, H3C 3J7 }
\affiliation{$^2$  Department of Physics and Astrophysics, University of Delhi, Delhi, India  110007}

\begin{abstract}
We study numerically the existence in a false vacuum, of magnetic monopoles which are ``thin-walled'', \ie, which correspond to a spherical region of radius $R$ that is essentially trivial surrounded by a wall of thickness $\Delta\ll R$, hence the name thin wall, and finally an exterior region that essentially corresponds to a pure Abelian magnetic monopole.  Such monopoles were dubbed false monopoles and can occur in non-abelian gauge theories where the symmetry-broken vacuum is actually the false vacuum. This idea was first proposed in \cite{Kumar:2010mv}, however, {the proof of the existence of thin-wall, false monopoles given there }was incorrect.  Here we fill this lacuna and demonstrate numerically{, for an appropriately modifed potential,} the existence of thin-wall false monopoles.  The decay via quantum tunnelling of the false monopoles could be of importance to cosmological scenarios which entertain epochs in which the universe is trapped in a symmetry broken false vacuum.
\end{abstract}
\maketitle
\section{Introduction}
Magnetic monopoles correspond to soliton solutions of non-abelian gauge field theories with scalar matter fields, where a simply connected simple gauge group symmetry is spontaneously broken to a smaller gauge symmetry where the gauge group has a $U(1)$ factor.  The topological possibility of the existence of monopoles requires that the homotopy group $\Pi_2(G/H)$ be non-trivial \cite{Goddard:1977da}.  It is well understood that, { from the short exact sequence $\Pi_2(G)\to\Pi_2(G/H)\to\Pi_1(H)\to\Pi_1(G)$, and the fact that $\Pi_2(G)=0$ and that $\Pi_1(G)=0$ implies } $\Pi_2(G/H)=\Pi_1(H)${.  But $H=U(1)$ and it is trivially known that $\Pi_1(U(1)) =\mathbb{Z}$, hence $\Pi_2(G/H)=\mathbb{Z}$,} and the integer characterizing the configuration is the monopole charge.  However, it can be that the symmetry breaking is temporary, that the symmetry breaking vacuum is a false vacuum that will eventually decay to the true vacuum.  Such a scenario adds new dynamical aspects to the monopole solution.  The universe can be trapped in this false vacuum phase for an interminably long period.  Indeed, cosmological solutions coming from string theory \cite{Kachru:2003aw} in fact propose that the present universe is in a false vacuum, that in principle has a very, very, long lifetime (hence the justification that it has survived for 13 billion years, the age of the universe).  However, topological solitons occurring in the false vacuum will promote the decay of the false vacuum, as we have shown in a series of papers \cite{Dupuis:2018utr,Dupuis:2017qyo,Haberichter:2015xga,Dupuis:2015fza,Lee:2013zca,Lee:2013ega,Kumar:2010mv}.  Precocious decay of the false vacuum would have undesirable consequences, if presumably, we are  living in one.

In this paper, we concentrate on the case of magnetic monopoles in a false vacuum.  In \cite{Kumar:2010mv} it was analyzed how thin wall monopoles could arise in the gauge field/Higgs field dynamics.  However, the analysis presented there relied on the proposition that the gauge field dynamics would essentially follow the Higgs field dynamics, which was not rigorously established, and seems not to be the case.  In this paper, we analyze numerically the magnetic monopole solutions that exist in a false vacuum.  The main difference is that the core of the monopole is in the true vacuum, and hence provides an outward pressure causing the monopole to be thin-walled.  Classically the monopole is stable, however, it can tunnel quantum mechanically \cite{paranjape_2017} to a classically unstable state, essentially where the radius of the core is large enough so that the core will expand uncontrollably, converting false vacuum in the exterior to the true vacuum in the interior.  This tunnelling amplitude was computed in  \cite{Kumar:2010mv}  the computation which relied only on the requirement that the monopole be thin-walled.  

The analysis in  \cite{Kumar:2010mv} assumed that the dynamics of the gauge field simply followed the dynamics of the Higgs field.  This does not seem to be correct, and hence the proof of the existence of thin-walled monopoles given there is not accurate.  In this paper, we perform the required numerical analysis to confirm the existence of thin-walled monopoles for an appropriate choice of the scalar field potential.  
\maketitle

\section{False magnetic monopoles and their tunnelling decay}
The monopole is constructed with a variation of the standard 'tHooft-Polyakov monopole where the field content consists of a triplet scalar $(\vec\phi)_i$, $i=1,2,3$ and the corresponding gauge field $\vec A_\mu$  of local $SO(3)$ gauge invariance.  
The potential for the scalar field, as treated in \cite{Kumar:2010mv},  was a 6th order potential, writing $h=|\vec\phi|$
\begin{equation}
V(\phi) = \lambda h^2 (h^2 - a^2)^2 + \gamma^2 h^2 -\epsilon .
\label{po}
\end{equation} 

The potential has the form given in Fig.\eqref{po},  and comprises of a true vacuum at $h=0$ and a false vacuum of the symmetry broken type at $h=\eta$ where
\begin{equation}
\eta = \sqrt{\frac{2{a}^2}{3} + \frac{\sqrt{{a}^4{\lambda}^2 - 
3{\gamma}^2{\lambda}}}{3{\lambda}}}.
\end{equation}   
The parameters of the potential are chosen so that the energy density difference is $\epsilon$ which normalizes the potential so that the false vacuum has energy density zero, evidently $V(h=0)=-\epsilon$.  Our idea was to mimic the thin-wall instanton analysis done by Coleman \cite{Coleman:1977py} in his study of false vacuum decay via the tunnelling formation of thin-walled bubbles of true vacuum, within the false vacuum.  

The spherically symmetric ansatz is:
\begin{eqnarray}
\phi_a &=& \hat{r}_a \, h(r) \nonumber \\
A_\mu^a &=& \epsilon_{\mu ab}\,\hat{r}_b \,\frac{1 - K(r)}{er} \nonumber \\
A_0 &=& 0\label{ansatz}
\end{eqnarray} 
and the equations of motion giving rise to the monopole solution are given by:
\begin{eqnarray}
h'' + \frac{2}{r} h' - \frac{2h}{r^2}K^2 - \frac{\partial V}{\partial h} &=& 0 \label{eomh} \\
K'' - \frac{K}{r^2}(K^2 - 1) - e^2h^2K &=& 0.
\label{eomk}
\end{eqnarray} 
For the original 'tHooft-Polyakov monopole configuration \cite{Hooft:1974qc,Polyakov:1974ek}, the potential has a maximum at $h=0$ and a symmetry-breaking minimum at $h = a\ne 0$, which is the global minimum of the potential.  For the false monopoles that we are considering, the potential has a global minimum at $h=0$ which does not break the symmetry and is the true vacuum, and has a local minimum at $h=\eta\ne0$ which corresponds to the false vacuum.  

We imagine that the universe is trapped in the false vacuum, which is a standard scenario\cite{Turner:1982jl}.   However, as the vacuum is not given by a single value of the scalar field but in fact a whole manifold of values, the manner in which the universe is trapped could vary from spatial point to point.  This allows for the possibility of topologically non-trivial configurations, within the false vacuum.  We have dubbed such monopoles ``false monopoles'' and they are seen, through our analysis \cite{Kumar:2010mv}, to be unstable.  The 6th order potential \eqref{po} with the appropriate choice of parameters, does correspond to a potential with a true minimum at $h=0$ and a false minimum at $h=\eta$ where $\epsilon$, the difference of energy density, can be taken to be arbitrarily small.   

The thin-walled false monopole solution that we are looking for would be classically stable.  We informally divide space into the interior and the exterior of the monopole.  The exterior is where the fields are essentially abelian, while the interior of the monopole contains the full non-abelian structure.  The topological twisting of the scalar field in the exterior region at spatial infinity requires that the scalar field must vanish in the interior of the monopole for the configuration to be non-singular.  Hence there is a region in the interior where the configuration is essentially in the true vacuum.  This region would naturally be unstable to expand, as it has negative energy density, $-\epsilon$, compared to the false vacuum in the exterior which has been normalized to have vanishing energy density.  For a solution of the thin-wall type, this expansion is arrested because the energy in the wall increases as the radius squared, while the negative energy increases as the radius cubed.  For small radius, the positive contribution to the energy dominates, but eventually, for large radius, it is clear that the negative contribution to the energy is dominant.  The size of the interior region is stable under collapse due to the Coulomb energy of the monopole.  The energy in the abelian magnetic field will diverge inversely with the radius of the interior region consequently arresting the collapse.  The eventual instability of the false monopole is due to quantum tunnelling.   The radius of the bubble can tunnel out to a larger radius at which the negative energy density of the true vacuum becomes dominant and the monopole size expands without bound, converting false vacuum into true vacuum. 

\begin{figure}[!htp]
\begin{center}
\includegraphics[width=.45\textwidth]{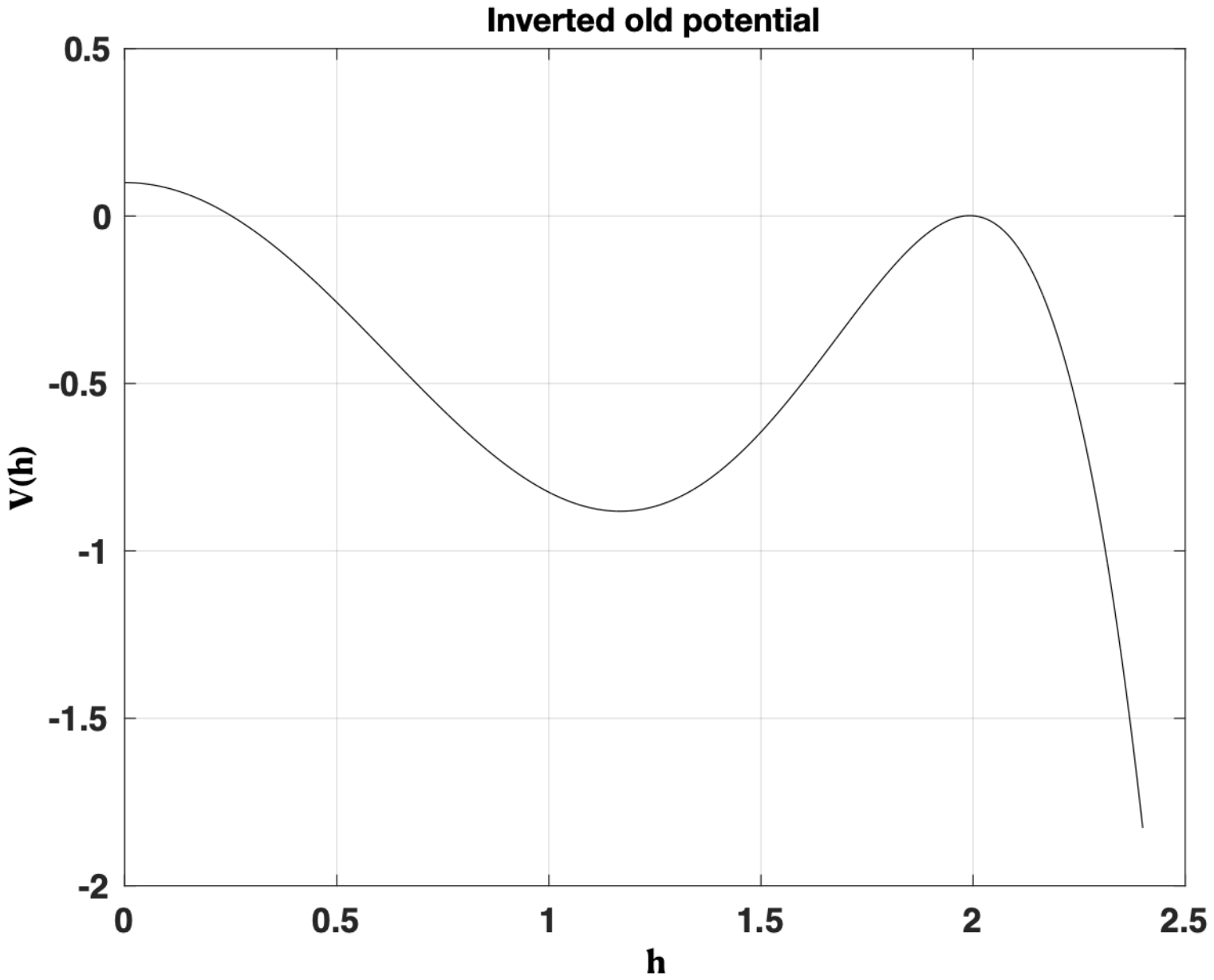}
\caption{Inverted potential of Eqn.\eqref{po} for the scalar field at generic values of the parameters.}
\label{fop}
\end{center}
\end{figure}

The equations of motion \eqref{eomh} and \eqref{eomk} can be interpreted as the equations governing the position of two particles, $h$ and $K$, as a function of ``time'' corresponding to $r$.  The $h$ dynamics correspond to movement in the presence of the inverted potential $-V$, as depicted in Fig.\eqref{fop}, but also in the presence of a positive impulse due to the $K$ dependent term $2 hK^2/r^2$ and a ``time'' dependent ``drag'' term $-2h'/r$.  The drag term is called so because normally friction is included, in an otherwise conservative equation of motion, through just such a first derivative term.  Both of these non-standard terms disappear as $r$ becomes large.  The $K$ equation of motion does not directly depend on the potential, however, it does implicitly do so through its dependence on $h$.  Intuitively, we expect that the $h$ field will remain at or near $h=0$ if it starts very near $h=0$ or starts directly at $h=0$ with a very small initial velocity.  By changing the values of the parameters, we can adjust how long the field remains at $h$ nearly zero.  If $r$ becomes large, both the impulse term (that depends on the $K$ field)  and the drag term which depends on the derivative $h'$ become negligible, as they are multiplied $1/r^2$ and $1/r$ respectively.   The subsequent motion is conservative and can be easily analyzed.  

In \cite{Kumar:2010mv}, it was invoked that the $K$ evolution would simply follow the $h$ evolution, hence if the $h$ field stayed for a long time at $h=0$, then the $K$ field would stay at $K=1$ until the $h$ field started its non-trivial evolution.  This was apparently not true.  Numerical studies show that it is largely the boundary conditions for large $r$ that govern the behaviour of the $K$ field, and even though the $h$ field remains zero, the $K$ field begins its trajectory towards a vanishing value independently of the value of the $h$ field.  Furthermore, it was found that even though we can make the initial potential as flat as we like near $h=0$, it still does not give rise to a thin wall solution.  The $h$ field simply starts to move up to its asymptotic value $h=\eta$ from $r\approx 0$.  The reason for this behaviour has to do with the impulse term $\frac{2h}{r^2}K^2$ which pushes the $h$ field along immediately at $r=0$.  Therefore, the idea of obtaining a thin wall monopole with the potential Eqn.\eqref{po} was not correct.  The content of this paper is to show how to modify the potential in order to obtain a thin-wall, false monopole solution. 

\section{Modified potential}
The potential can be modified in an infinite number of ways, as we are not concerned with renormalizability.  { We are primarily motivated by the desire to exhibit a potential for which the false monopoles will present themselves as thin-wall monopoles.  However, we find that the modifications are justified with respect to current models of cosmology.  Our modification will give rise to a false vacuum that is not only classically stable, but also comprises of a very wide flat region.  Such false vacua with wide flat regions have been considered in many viable cosmological models.  For example, the KKLT solution for a cosmology coming from string theory, \cite{KKLT} obtains that the present universe is actually in a false vacuum phase where the very wide flat aspect of the potential gives that the tunnelling probability to the true vacuum is much longer than the present age of the universe.  In supersymmetric models, the existence of flat directions is generic and required, \cite{ENQVIST200399}.  However, non-perturbative effects and supersymmetry breaking can slightly lift degeneracies of supersymmetric vacua, giving rise to just the kind of very flat vacua that we will be modelling.   In the inflationary cosmology scenario, the notion of slow-roll inflation requires a false vacuum to be very flat such that the scalar field slowly rolls down a potential rather than tunnelling out of it, \cite{LINDE1982389,PhysRevLett.48.1220}.  Inflation occurs if the roll is slow compared to the expansion of the universe.  However, eventually, the roll speeds up, arrests the inflationary phase and gives rise to particle creation.  Hence these scenarios all require a very flat false vacuum, the kind of false vacuum that we will be studying. Indeed, if the disintegration of the false vacuum that we are investigating is in fact relevant in cosmological scenarios where the universe is in a false vacuum for an appreciable amount of time, it would require serious readjustments of those scenarios that rely on a long period in which the universe is trapped in such a false vacuum. }

\begin{figure}[!htp]
\begin{center}
\includegraphics[width=.45\textwidth]{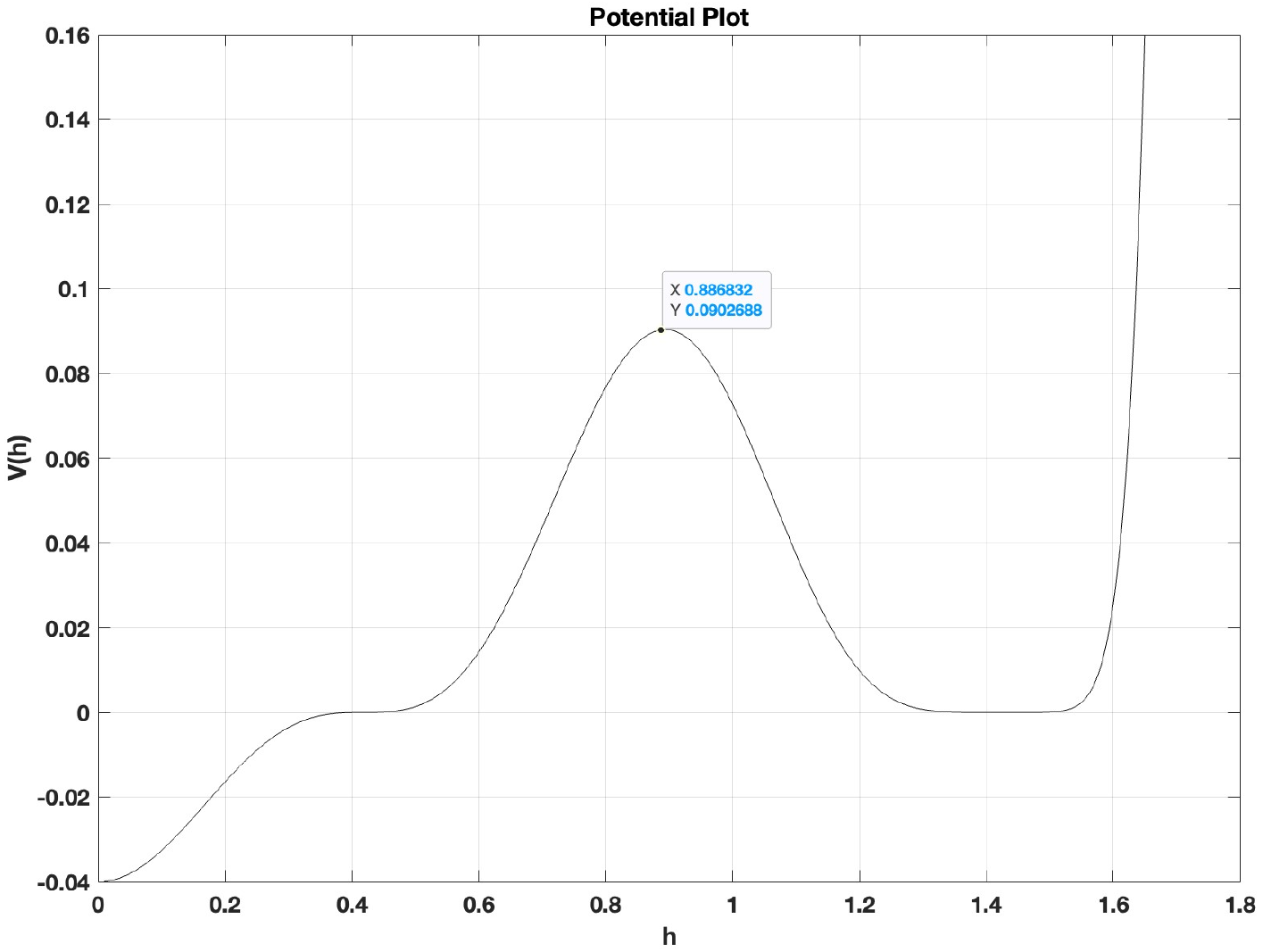}
\caption{Potential for the scalar field with $n=1$, $a=1.43$, $\lambda=0.1$ and $\epsilon=.4$}
\label{figpot}
\end{center}
\end{figure}

The potential that we use should simply be thought of as an effective potential.  The potential should have an absolute minimum at $h=0$ and additionally a local minimum (false vacuum) at $h=a$, the energy density difference between the minima should be an adjustable parameter.  The energy density of the false vacuum is normalized to zero.  We consider the following form for the potential,  
\beq
V(h)=\lambda\Big( (h^2-a^2)^2(h^2-\frac{\epsilon^{\frac{1}{2n+1}}}{a^4})\Big)^{2n+1},
\eeq
as shown in Fig\eqref{figpot}.
The potential is symmetric under $h\to -h$ however we are only interested in the range $h\ge 0$.  Here, the potential has two roots at $h_i= {\frac{\sqrt{\epsilon^{\frac{1}{2n+1}}}}{a^2}}$ and at $h= a$.  

For $\epsilon^{\frac{1}{2n+1}}<a^6$  the potential rises up from $-\lambda\epsilon$ at $h=0$ to an inflection point (and root) at $h_i$ followed by a maximum at $\bar h$ and then followed by a local minimum (and root) at $h=a$.  Afterwards, it rises up to $+\infty$.  This is depicted in Fig.\eqref{figpot}.  However, the three critical points can exchange their order.  

\begin{figure}[!htp]
\begin{center}
\includegraphics[width=.45\textwidth]{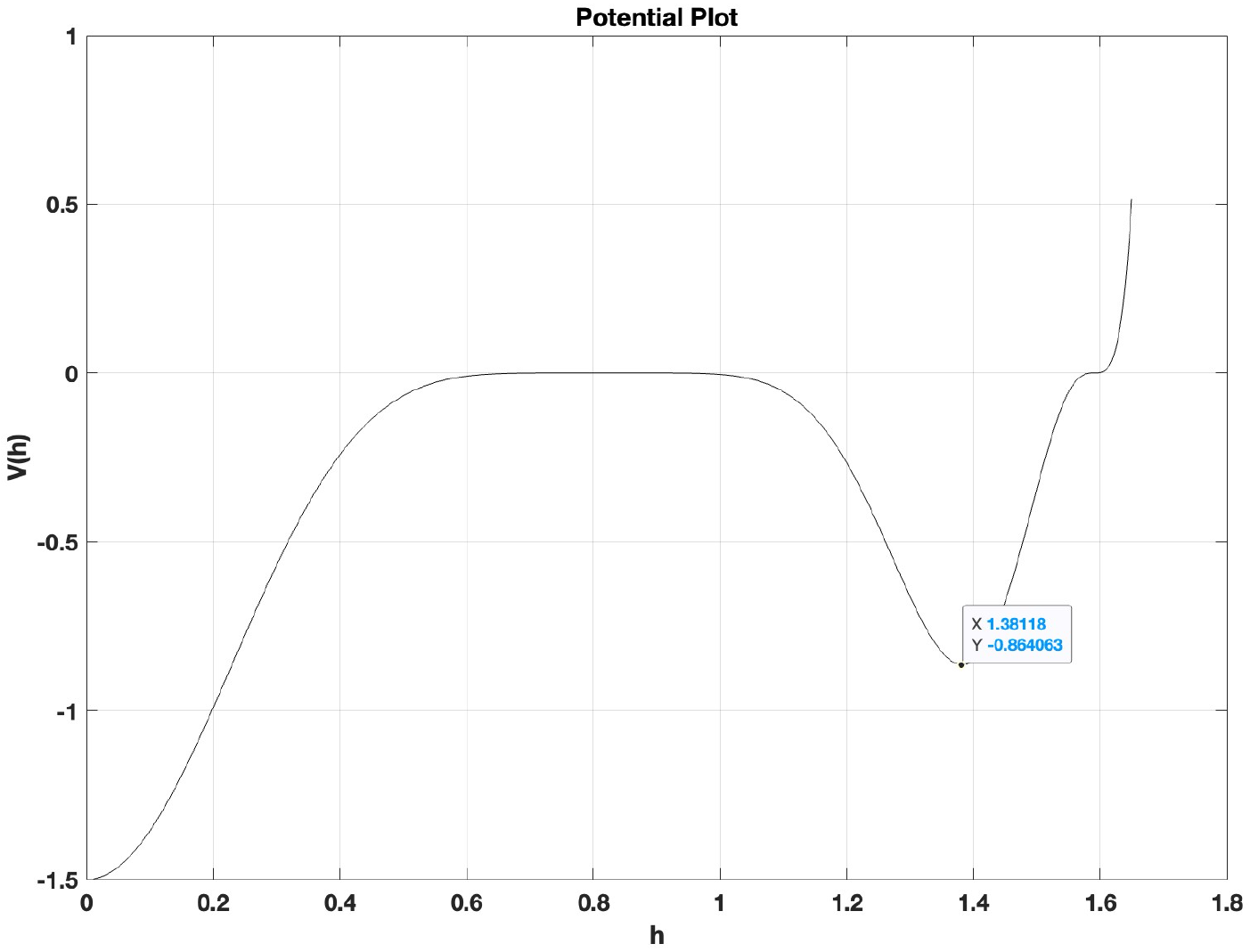}
\caption{Potential for the scalar field with $n=1$, $a=.82$, $\lambda=1$ and $\epsilon=1.5$}
\label{figpot2}
\end{center}
\end{figure}

For $\epsilon^{\frac{1}{2n+1}}>a^6$  we get the order interchanged. 
This is depicted in Fig.\eqref{figpot2}.  
We are not interested in a potential with this behaviour, 
therefore we will 
restrict ourselves to the region $\epsilon^{\frac{1}{2n+1}}<a^6$.

In order to confirm the described behaviour of the potential, we first compute its derivative.   
Writing $\bar\epsilon=\epsilon^{\frac{1}{2n+1}}$ we have
\bea
V'(h)&=&\lambda(2n+1)\Big( (h^2-a^2)^2(h^2-\frac{\bar\epsilon}{a^4})\Big)^{2n}\nonumber\\
&\times&\big((2(h^2-a^2)(h^2-\frac{\bar\epsilon}{a^4})+(h^2-a^2)^2)\big)2h \nonumber\\
&=&\lambda(2n+1) (h^2-a^2)^{4n+1}(h^2-\frac{\bar\epsilon}{a^4})^{2n}\nonumber\\
&\times&\big(3h^2-(\frac{2\bar\epsilon}{a^4}+a^2)\big)2h 
\eea
and one can read off the critical points $V'(h)=0$ at $0,\,\, h_i=\sqrt{\bar\epsilon}/a^2,\,\,\bar h=\sqrt{(2\bar\epsilon/a^4+a^2)/3}  $ and $a$.
\begin{enumerate}
\item
The global minimum at $h=0$ is simple with {$V(0)=-\lambda\epsilon$} and clearly $V''(0)=\lambda(2n+1) a^2\bar\epsilon^{2n}({2\bar\epsilon}/{a^4}+a^2)>0$.  

\item
The critical point at $ h_i$ is an inflection point and a root of order $2n+1$.  {We find}
\beq
\hskip2mm V^{2n+1}(h_i)=\lambda (2n+1)!(h_i^2-a^2)^{4n+2} (2h_i)^{2n+1}\nonumber
\eeq
which gives the leading contribution in the Taylor expansion of the potential as $\sim (h-h_i)^{2n+1}$.   This is an odd power $h-h_i$  changing sign on either side of $h_i$, implying an inflection point. 
\item
The critical point at $\bar h$ is a simple local maximum.  We find $V''(\bar h)$ by differentiating the term $3h^2-(\frac{2\bar\epsilon}{a^4}+a^2)$ in $V'(h)$ and setting $h=\bar h$ afterwards.  The contributions from differentiating the other terms of course vanish as the term vanishing at $h=\bar h$ is intact.  Thus we find 
\beq
\hskip8mm V''(\bar h)=-\lambda (2n+1)2^{4n+1}((a^2-\bar\epsilon/a^4)/3)^{6n+1} 12 \bar h.\nonumber
\eeq  
We observe that the sign of the first factor changes from positive to negative as $\bar\epsilon$ passes from below $a^6$ to above as the power $6n+1$ is odd.  The other factors do not change sign.  Hence for $\bar\epsilon< a^6$, the region that we are interested in, we have { $V''(\bar h)< 0$} \ie  a maximum at $\bar h$.  
\item
The critical point at $h=a$ is also a $4n+2$ order root, \ie\,  $V(h)\sim (h-a)^{(4n+2)}$ with first non-vanishing derivative is of even order, $V^{(4n+2)}(a)\ne 0$.  Differentiating $V(h)$, $4n+2$ times and then evaluating at $h=a$ only gives a non-vanishing contribution when the derivative acts on the $h-a$ factor.   We easily find $V^{4n+2}(a)=\lambda(4n+2)! (2a)^{4n+2}(a^2-\bar\epsilon/a^4)^{2n+1}$ which, if $\bar\epsilon<a^6$, is positive signifying a local minimum.  
\end{enumerate}

\section{Numerical results}
The equations of motion Eqns.\eqref{eomh} and \eqref{eomk} correspond to dynamics in minus the potential, and as if $r$ is the time.  We have experimented with various values of $\epsilon$, $a$ and $n$. We are able to find thin-wall type (false) monopole solutions.

\begin{figure}[!htp]
\begin{center}
\includegraphics[width=0.45\textwidth]{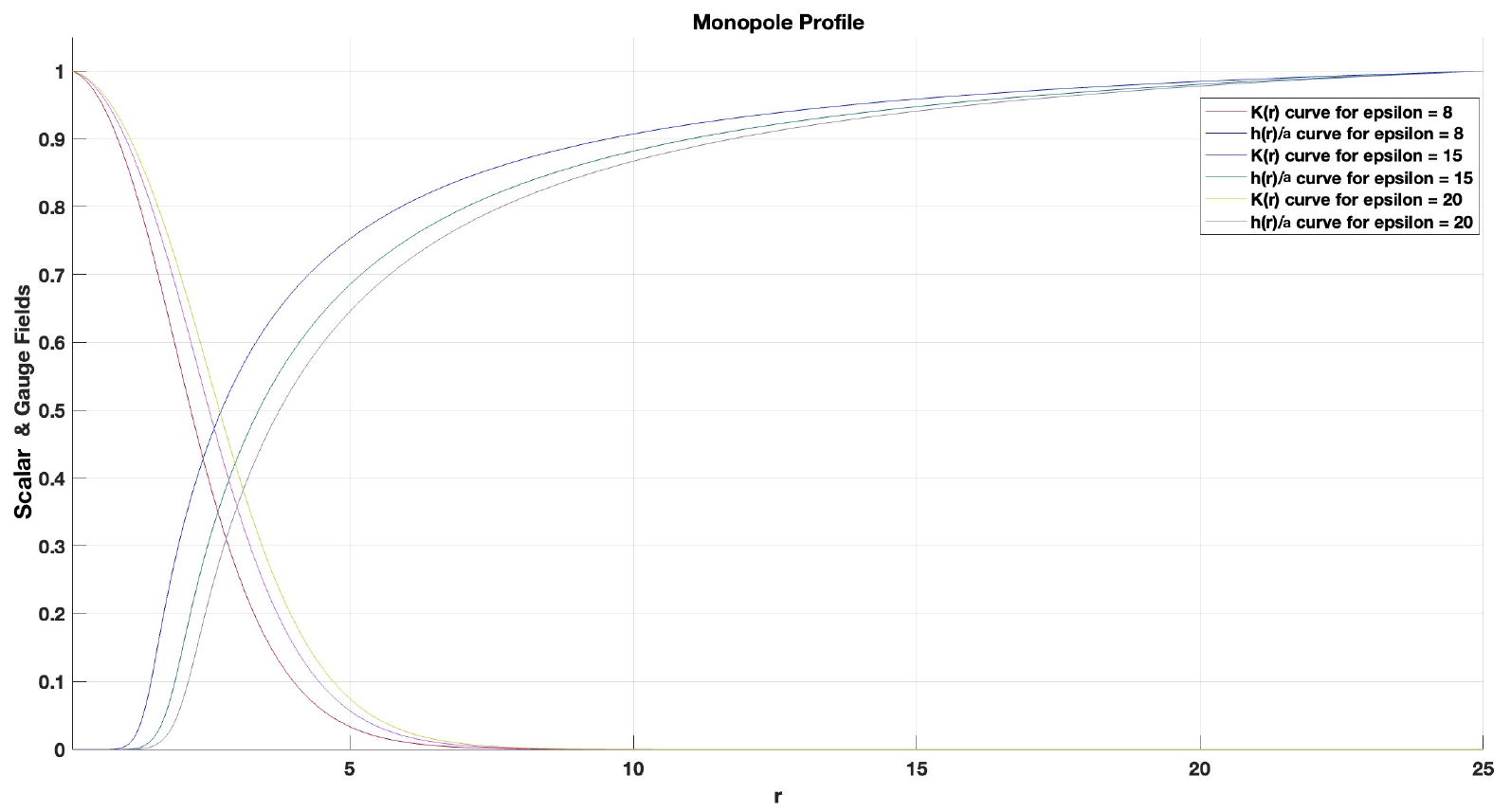}
\caption{The profiles for $h/a$ and $K$ for various values of $\epsilon$ {and for $\lambda=0.1$, $a=1.4$ and }$n=4$.}
\label{twp}
\end{center}
\end{figure}

\begin{figure}[!htp]
\begin{center}
\includegraphics[width=0.45\textwidth]{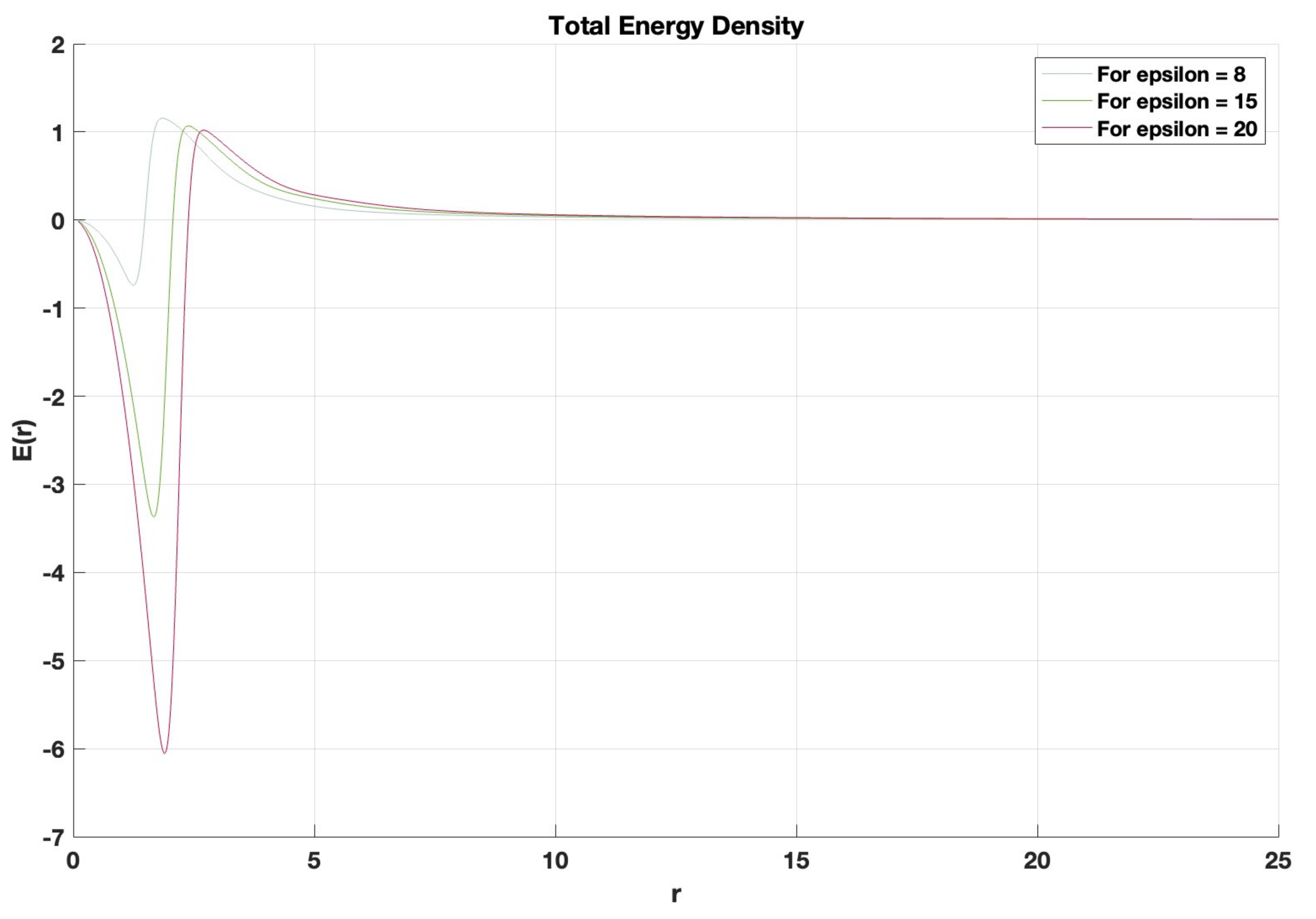}
\caption{The total energy density for the profiles in Fig.\eqref{twp}for various values of $\epsilon$ {and for $\lambda=0.1$, $a=1.4$ and }and for $n=4$.}
\label{ted}
\end{center}
\end{figure}

\begin{figure}[!htp]
\begin{center}
\includegraphics[width=0.45\textwidth]{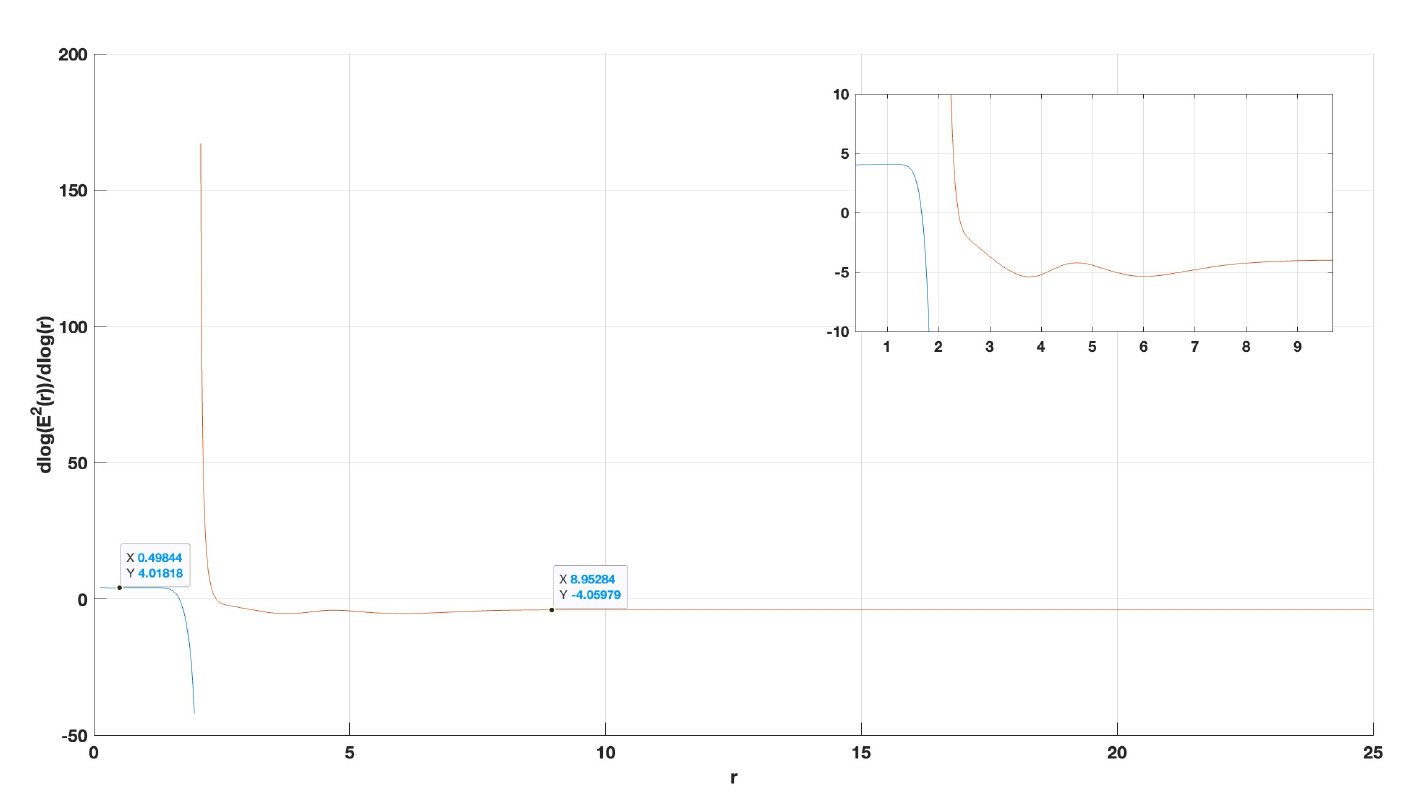}
\caption{The derivative of the log of the total energy density squared with respect to the log of $r$, for $\epsilon=15$ {and for $\lambda=0.1$, $a=1.4$ and } and  $n=4$.}
\label{le2}
\end{center}
\end{figure}

The work is numerical, using a Matlab code to find the solution, which is included in the supplementary materials.  We find the monopole profiles have a thin-wall behaviour as defined by the energy density of the configuration.  The field profiles for $h$ and $K$ are given in the Figs. \eqref{twp}.  
These profiles correspond to thin-wall solutions when we look at the total energy density shown in Fig.\eqref{ted}.  
The total energy density starts at zero and descends as $\sim -r^2$, which we confirm numerically.  This is expected {as} the potential energy in the scalar field is $-\lambda\epsilon$ when $h\approx 0$.  Then  {the energy density} should behave simply as $-\lambda\epsilon r^2$ {including} the contribution of the volume element.  There is a competing positive contribution from the magnetic field energy density as $K$ starts to descend from $K=1$, however, in the interior, the total energy density  {seems to be} dominated by the negative scalar potential energy.  The descent of the total energy density is arrested when the scalar field begins to move away from $h=0$.  This is where the wall starts, in Fig.\eqref{ted} and implicitly from the profiles in Fig.\eqref{twp}  we see that the increase in the energy density is rather brusque.  This allows for a well-defined value for the inner radius of the wall.  The outer radius of the wall starts once the total energy density begins to get its dominant contribution from the Coulomb energy of the exterior abelian magnetic field.  This magnetic field is constructed from the non-abelian magnetic field and the covariant derivatives of the scalar field. The contribution to the total energy density from the scalar field potential becomes negligible as the scalar field assumes its value at the false vacuum (which has been normalized to zero energy density) in the exterior.  The magnetic field behaves as $\sim 1/r^2$, hence the total energy density behaves as $\sim r^2 (1/r^2)^2=1/r^2$.  To exhibit this numerically, we consider the log of the total energy density squared, $\ln (E(r))^2$.  We take the square so that the log acts on a positive function.  The slope of this function should behave as $\approx 4$ in the interior region and as $\approx -4$ in the exterior region, with a complicated interpolation between these values within the wall region.  In Fig.\eqref{le2}, we plot the derivative (with respect to $\ln r$) of the log of the total energy density squared.  We see clearly that the interior value is indeed about +4 while the exterior value is -4 as expected.  Additionally, we see that the wall thickness $\Delta$, over which the behaviour of the energy density interpolates from $\sim -r^2$ to $\sim 1/r^2$, is relatively small compared to the wall radius.  

\begin{figure}[!htp]
\begin{center}
\includegraphics[width=0.45\textwidth]{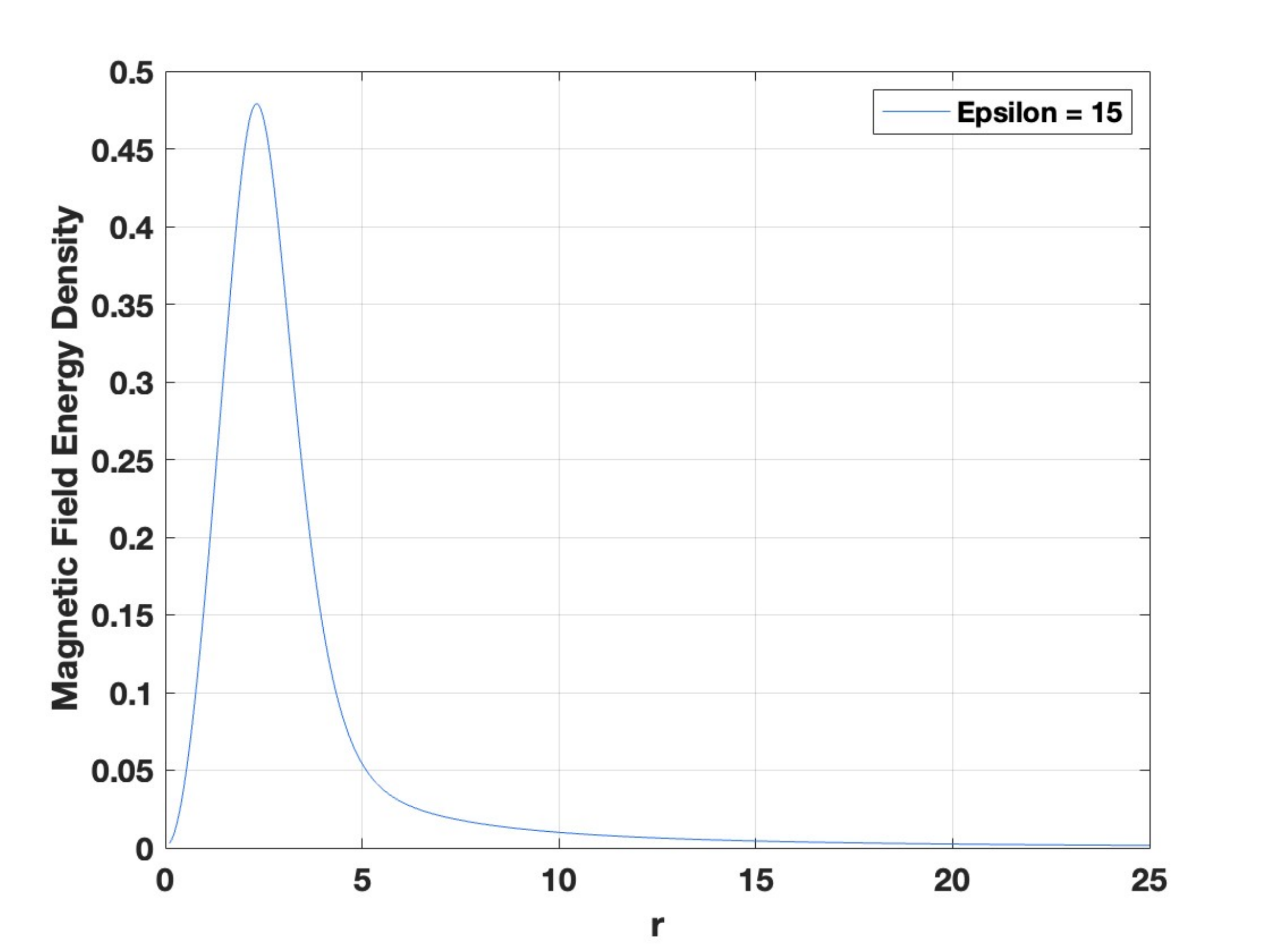}
\caption{The magnetic energy density for $\epsilon=15$ {and for $\lambda=0.1$, $a=1.4$ and } and  $n=4$.}
\label{med}
\end{center}
\end{figure}

\begin{figure}[!htp]
\begin{center}
\includegraphics[width=0.45\textwidth]{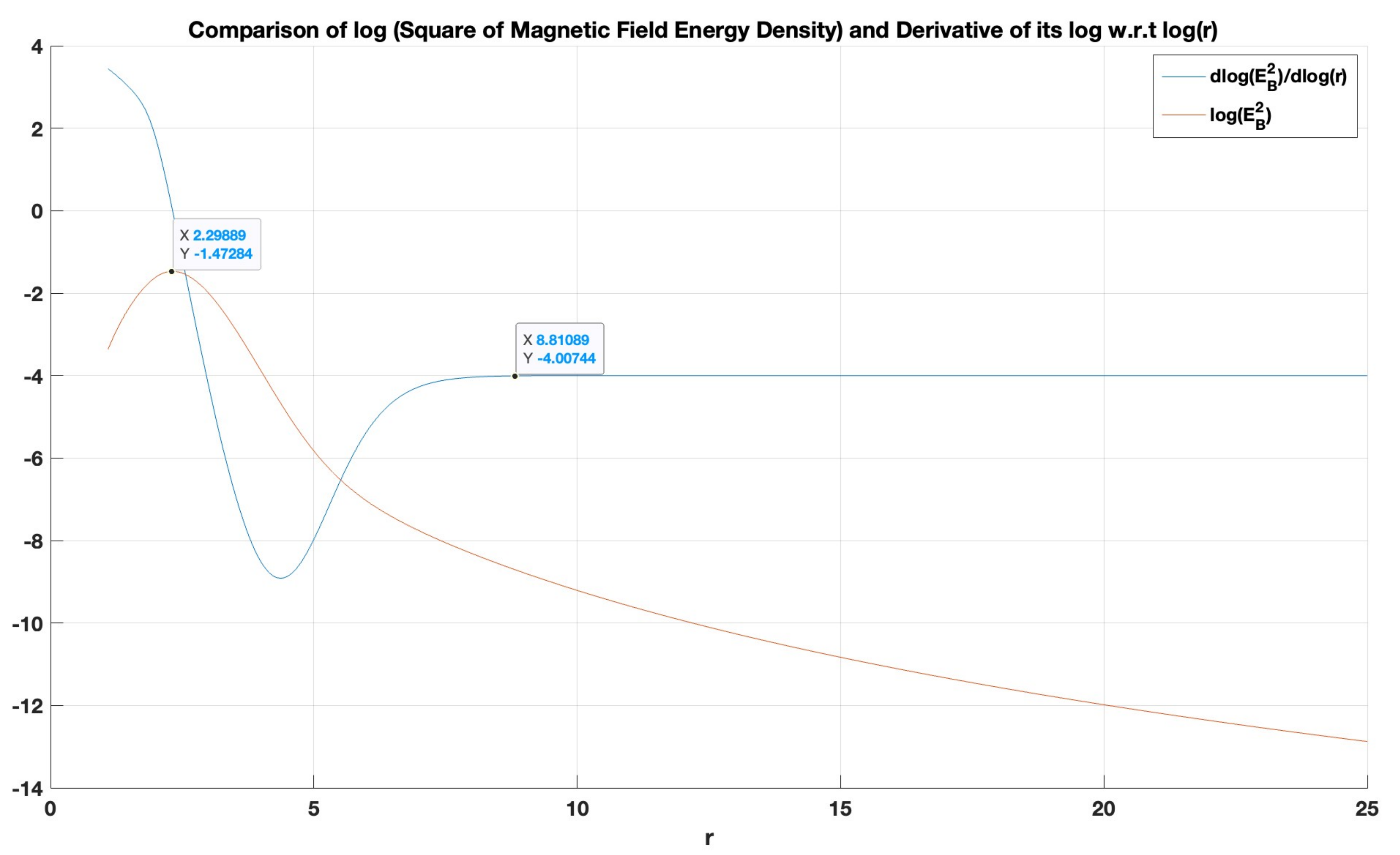}
\caption{The log of the magnetic energy density squared and its derivative (by $\ln r$) for $\epsilon=15$ {and for $\lambda=0.1$, $a=1.4$ and } and  $n=4$.}
\label{dmedsq}
\end{center}
\end{figure}

The non-abelian magnetic energy density behaves in a complicated way in the interior and as a $1/r^2$ contribution in the exterior, as can be seen from Fig.\eqref{med} and Fig.\eqref{dmedsq}.  The magnetic field energy density rises up to a peak at the wall and then descends downwards to the expected $\sim 1/r^2$ behaviour from its Coulomb tail.  The derivative by $\ln r$ of the log of the magnetic field energy density squared asymptotes to -4 confirming that the magnetic field energy density behaves like $\sim 1/r^2$ in the exterior.

\begin{figure}[!htp]
\begin{center}
\includegraphics[width=0.45\textwidth]{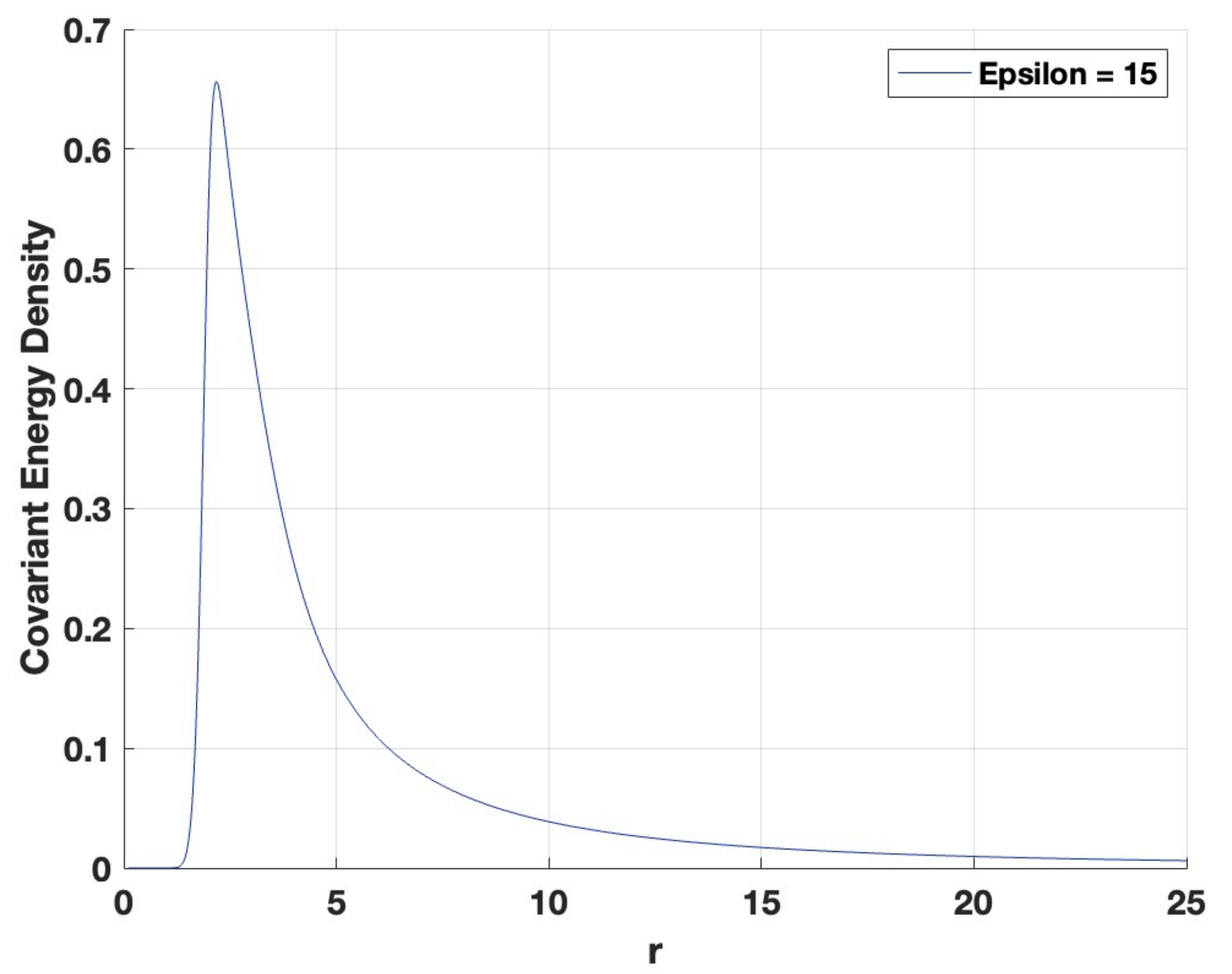}
\caption{The covariant derivative energy density for $\epsilon=15$ {and for $\lambda=0.1$, $a=1.4$ and } and  $n=4$.}
\label{covdev}
\end{center}
\end{figure}

\begin{figure}[!htp]
\begin{center}
\includegraphics[width=0.45\textwidth]{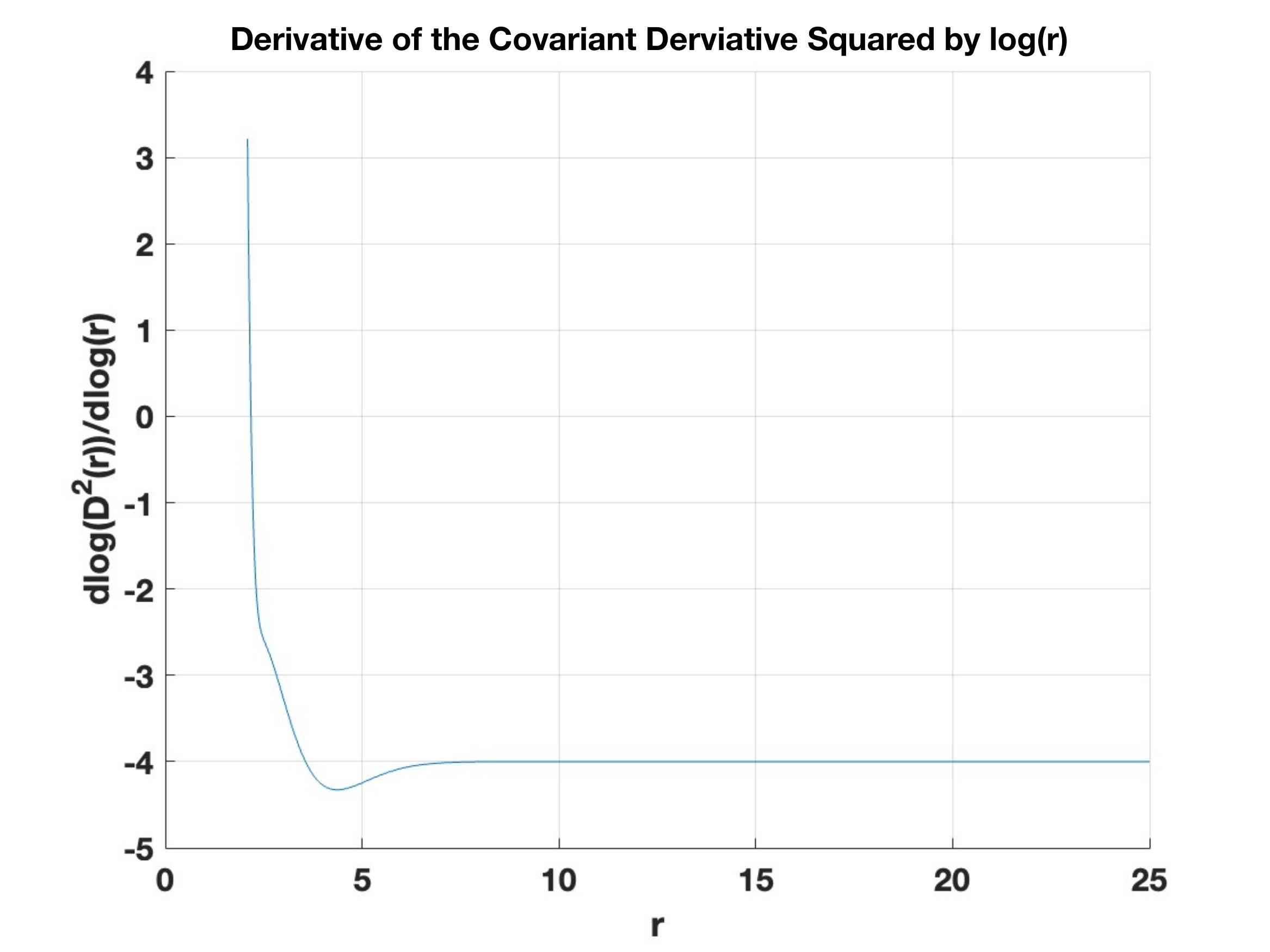}
\caption{The derivative by $\ln r$ of the log of the covariant derivative energy density squared for $\epsilon=15$ {and for $\lambda=0.1$, $a=1.4$ and } and  $n=4$.}
\label{cdd}
\end{center}
\end{figure}

In Fig.\eqref{covdev} we plot the energy density in the covariant derivative of the scalar field and in Fig.\eqref{cdd} the derivative by $\ln r$ of the log of its contribution to the energy density squared.  We observe a vanishing contribution in the interior, then a brusque rise up at the inner wall radius, and then an interpolation to the exterior energy density contribution which behaves exactly like $\sim 1/r^2$ as seen from the exterior behaviour in Fig.\eqref{cdd} which assumes the corresponding asymptotic value of -4.

\begin{figure}[!htp]
\begin{center}
\includegraphics[width=0.45\textwidth]{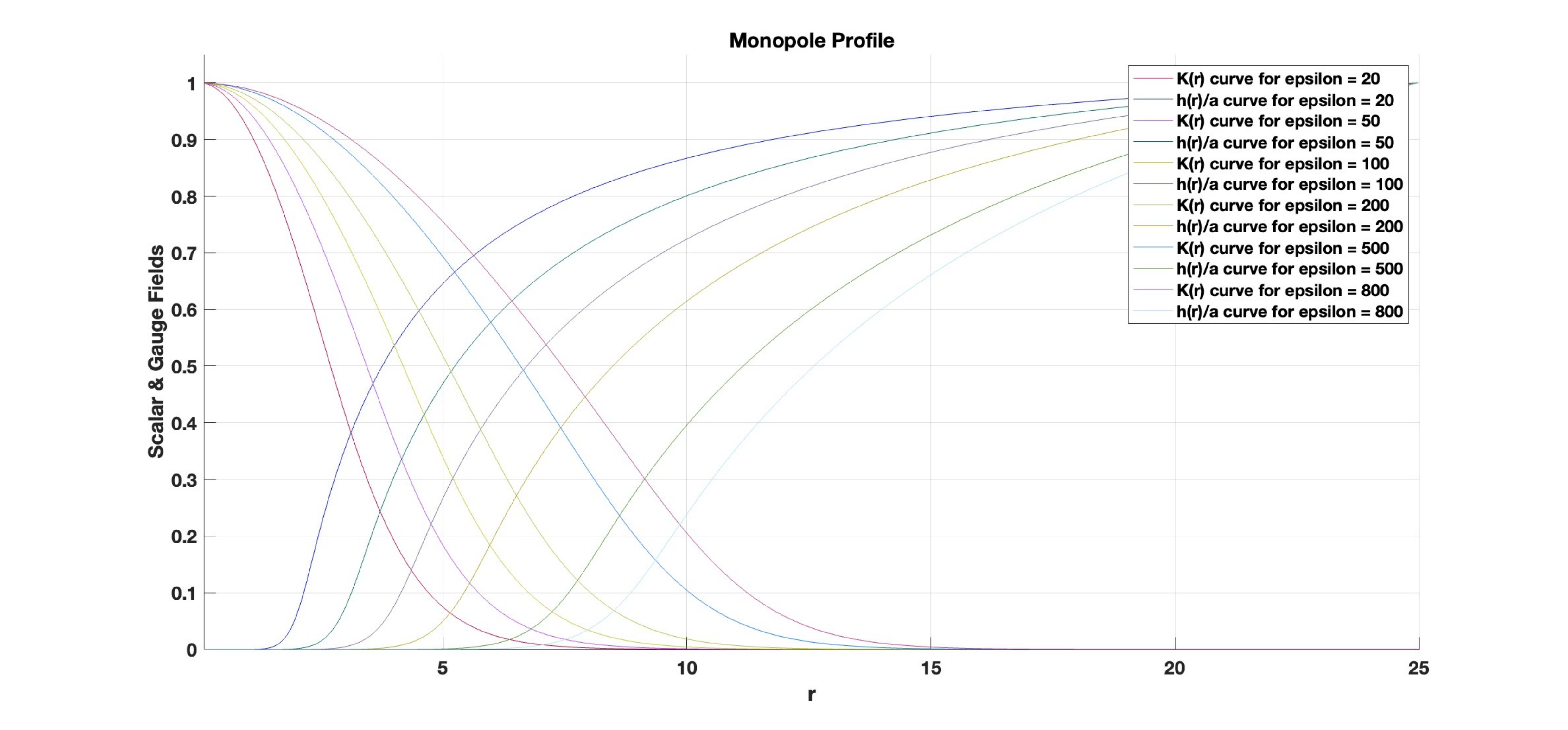}
\caption{Multiple solutions for $h$ and$K$ for various values of $\epsilon$ {and for $\lambda=0.1$, $a=1.4$ and } and  $n=4$.}
\label{mp}
\end{center}
\end{figure}

\begin{figure}[!htp]
\begin{center}
\includegraphics[width=0.45\textwidth]{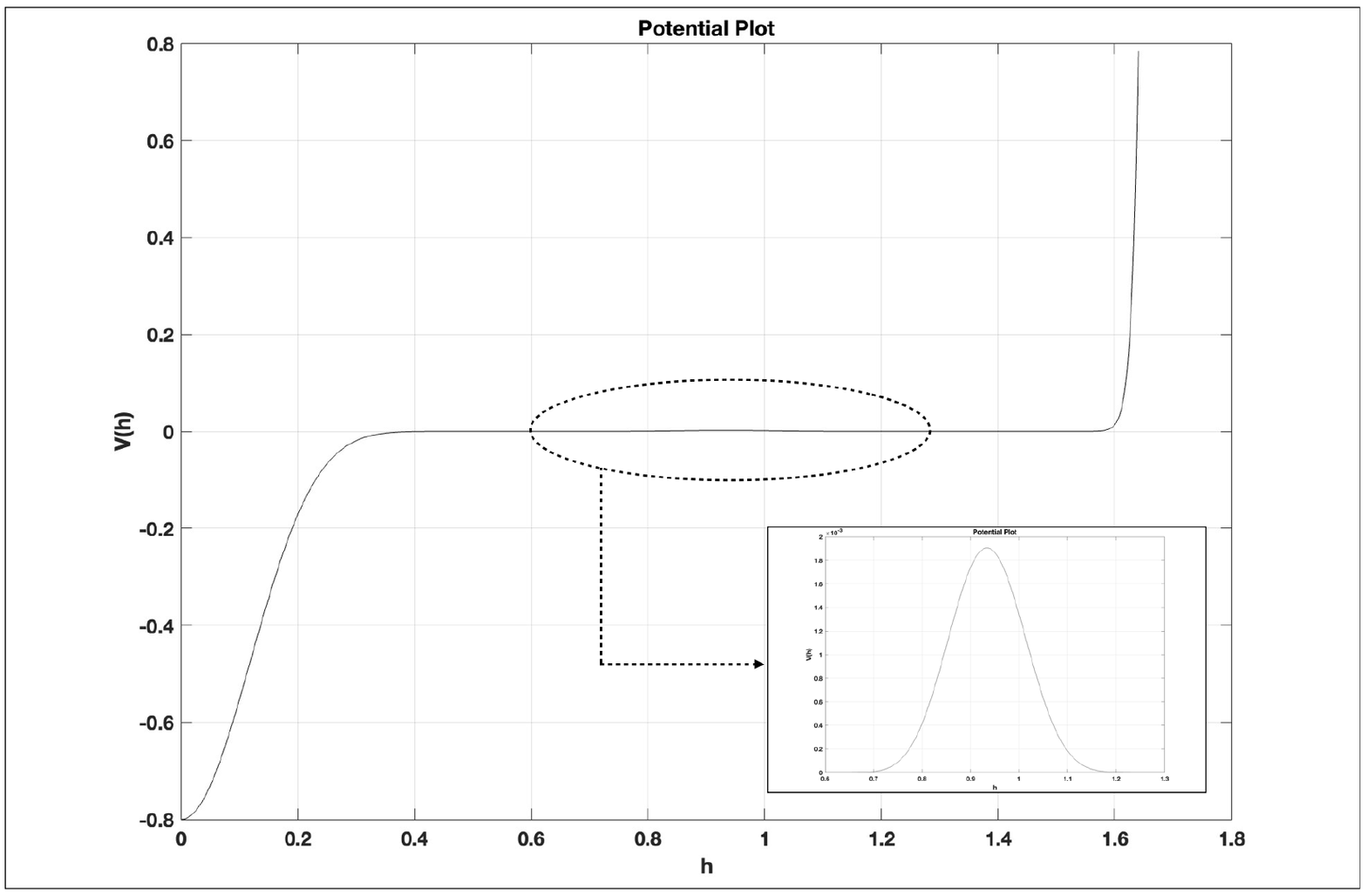}
\caption{ { Graph of the potential (including a zoom around $r=1$), for $\epsilon=8$, $\lambda=0.1$, $a=1.4$ and  $n=4$.}}
\label{pot8}
\end{center}
\end{figure}

\begin{figure}[!htp]
\begin{center}
\includegraphics[width=0.45\textwidth]{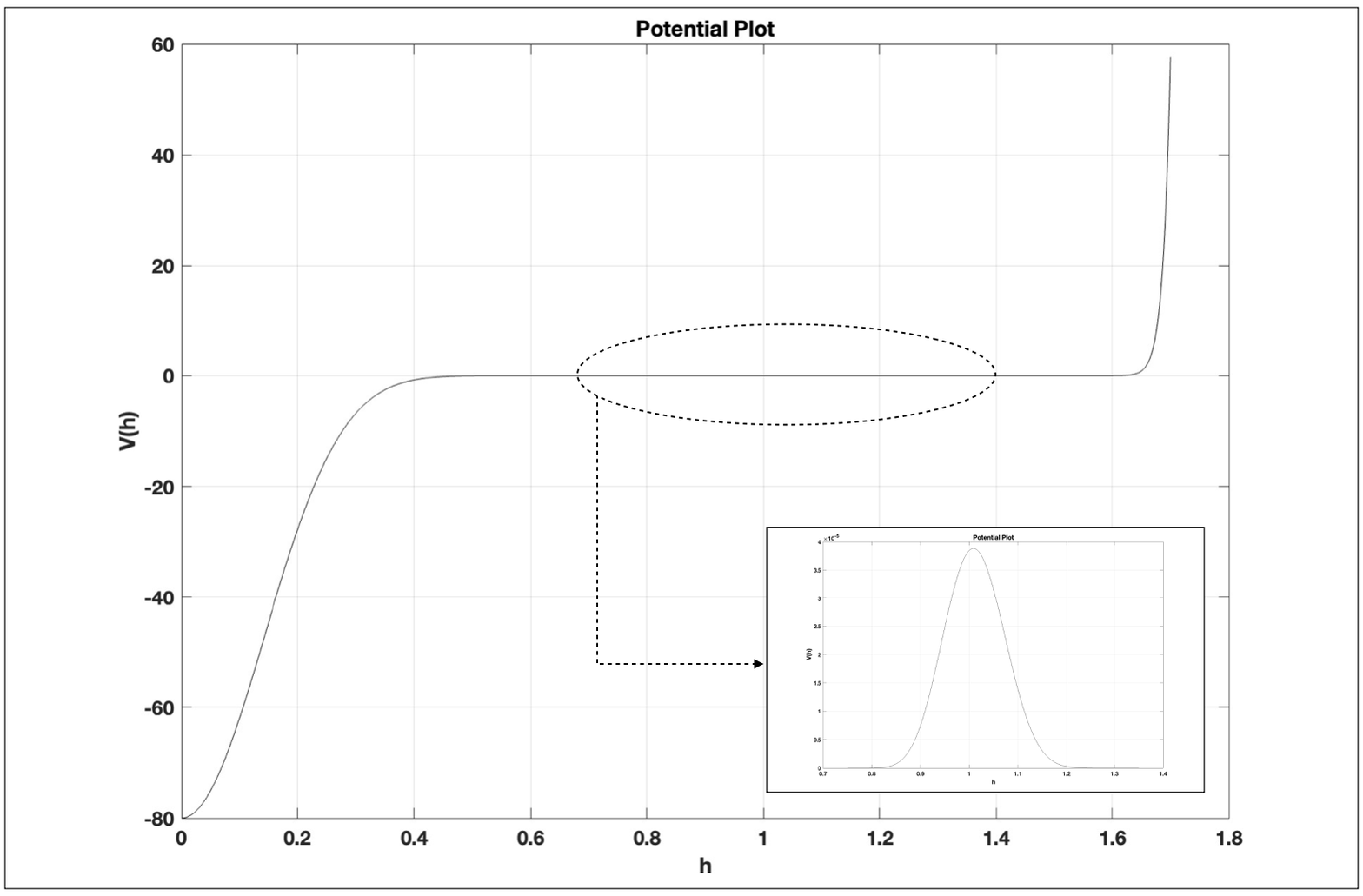}
\caption{{  Graph of the potential for $\epsilon=800$ (including a zoom around $r=1$), $\lambda=0.1$, $a=1.4$ and  and  $n=4$.}}
\label{pot800}
\end{center}
\end{figure}

We end this section with Fig.\eqref{mp} of multiple profiles for the solutions for $h$ and $K$ for various values of $\epsilon$.  { All the values of the parameters satisfy the condition that $\bar\epsilon< a^6$ as is required.  The values of $n$ range from $\epsilon=8$ in Fig. \eqref{twp}, to $\epsilon=800$ in Figure \eqref{mp}.  Explicitly for the values at the two extremes, we have $\epsilon=8$, $\lambda=0.1$ $a=1.4$ and $n=4$ giving $\bar\epsilon=1.26<7.53=a^6$ and for $\epsilon=800$, $\lambda=0.1$ $a=1.4$ and $n=4$ giving $\bar\epsilon=2.1<7.53=a^6$.}
Fig.\eqref{mp} gives the monopole profiles of $h$ and $K$ for 6 different values of $\epsilon$ and fixed values of $\lambda=0.1$, $a=1.4$ and $n=4$.  {It could be clarifying to see the potential explicitly that gives rise to the thin-walled monopoles.  Here, in Figs. (\ref{pot8},\ref{pot800}) we give the graph of the potential for the two extreme values, $\epsilon=8$ and $\epsilon=800$.  }
 
 
\section{Conclusions}
We will make several observations.  First, the idea that $\epsilon$ should be small is not the appropriate criterion, it seems.  Indeed, we get thin-wall type solutions as we take $\epsilon$ quite large, as can be seen in Fig.\eqref{mp}, although we always remain in the region $\bar\epsilon<a^6$.  Second, the thin-wall nature of the solution is quite different than in the case of thin-wall instantons that give rise to vacuum bubbles \cite{Coleman:1977py}.  With the vacuum bubbles, the interior of the bubble was a quiescent true vacuum.  In our case, the energy density of the true vacuum is dominant and normalized to be negative, however, the gauge field does give a non-zero contribution.  This is reminiscent of the false vortices and cosmic strings that have been studied in a similar context \cite{Lee:2013ega,Dupuis:2017qyo} where the interior contains non-trivial magnetic fields.  The profiles show an energy density that behaves as $\sim r^2$ in the interior.  Here the scalar field remains in the true vacuum, $h\approx 0$ until it starts its trajectory up to the false vacuum, where the wall nominally begins.  The gauge field however immediately begins its descent from $K=1$ giving a contribution to the energy density which behaves as $ \le r^2$.  However, the negative energy density of the true vacuum of the scalar field gives a contribution to the energy density that behaves as $\sim - \lambda\epsilon r^2$ and it dominates over the contribution to the energy density from the gauge fields, for $\epsilon$ sufficiently large.  
The wall region has a sharply defined inner radius, essentially where the scalar field begins to move away from $h=0$, but the exact outer boundary is not as sharp.  However, eventually, in the exterior, the behaviour of the energy density is that of the Coulomb energy of the abelian magnetic field. 

With this understanding of the existence of the thin-wall false monopoles, we can imagine that the further analysis done in \cite{Kumar:2010mv} is perfectly justified.  The interior of the monopole will give a contribution to the total energy that behaves as three terms, the interior negative energy proportional to the volume, the positive wall energy proportional to the area and the exterior Coulomb energy proportional to the inverse of the radius.  
\beq
E(R)=-\alpha R^3 +4\pi\sigma R^2 +\frac{C}{R}
\eeq
where $\alpha$ $\sigma$ and $C$ are calculable parmeters.  The function $E(R)$ descends from $+\infty$ at $R=0$ to $-\infty$ at $R=+\infty$ but it will have a classically stable minimum corresponding to a thin-wall false monopole as long its derivative has two positive critical points.  The first critical point corresponds to a local minimum which gives the radius of the false thin-wall monopole, while the second corresponds to the position at the height of the barrier under which the wall must tunnel to render the false vacuum unstable.   $E'(R)=0$ at the critical points which gives the equation
$-3\alpha R^2+8\pi\sigma R -\frac{C}{R^2}=0$.
Multiplying by $R^2$ and dividing through by $-C$ yields the quartic
\beq
\frac{3\alpha}{C} R^4-\frac{8\pi\sigma}{C} R^3 +1=0.
\eeq
For a small $C$, which is reasonable for small gauge coupling, the equation is equivalent to
\beq
A R^4 -B R^3 +1=0
\eeq
with $A,B\gg 1$.  This equation is easily, analytically solved by symbolic algebraic manipulation software (Mathematica) and yields 4 solutions.  A clear discriminant must be positive to have any real solution
\beq
-256 A^3 +27 B^4\ge 0
\eeq
\ie 
\beq
C\le \frac{(4\pi\sigma)^3}{16 \alpha^3}.\label{eql}
\eeq
Furthermore, if Eqn.\eqref{eql} is satisfied, it can be easily seen that in the limit $\frac{8\pi\sigma}{C},\,\, \frac{3\alpha}{C}\gg 1$, there are always two real positive solutions.

\section{ACKNOWLEDGEMENTS}
We thank NSERC, Canada for financial support and Nick Manton, Richard MacKenzie and Urjit Yajnik for useful discussions and critical comments.


\bibliographystyle{apsrev}
\bibliography{ref}

\end{document}